\documentclass[a4paper,fleqn,usenatbib]{mnras}

\usepackage{newtxtext,newtxmath}

\usepackage[T1]{fontenc}
\usepackage{ae,aecompl}
\usepackage{ulem}

\usepackage{graphicx}	
\usepackage{amsmath}	
\usepackage{amssymb}	

\usepackage{pdflscape}
\usepackage{multirow}
\graphicspath{
    {fig/}
}
\defcitealias{2015MNRAS.446.3943M}{M15}

\newcommand{\hi}{{\rm H}{\textsc i }}

\newcommand{\msun}{\hbox{$M_{\sun}$}}

\def\fhi{f_{\rm{HI}}}
\def\fhit{f_{\rm{HI,total}}}
\def\fhib{f_{\rm{HI,bulge}}}
\def\fhid{f_{\rm{HI,disc}}}

\def\mt{{M_{\rm \star,total}}}
\def\md{{M_{\rm \star,disc}}}
\def\mb{{M_{\rm \star,bulge}}}
\def\mhi{{M_{\rm HI}}}

\def\grb{{(g-r)_{\rm bulge}}}
\def\grd{{(g-r)_{\rm disc}}}

\title[Scaling relations of galactic bulge, disc and HI]{The growth of bulges and discs in relatively HI-rich galaxies: indication from HI scaling relations}

\author[Xinkai Chen et al.]{Xinkai Chen,$^{1,2}$\thanks{E-mail: cxk@mail.ustc.edu.cn}
Jing Wang,$^{3}$\thanks{Corresponding author. E-mail: jwang\_astro@pku.edu.cn}
Xu Kong,$^{1,2}$
Barbara Catinella,$^{4,5}$
Li Shao,$^{6}$
\newauthor  and Houjun Mo$^{7,8}$
\\
$^{1}$CAS Key Laboratory for Research in Galaxies and Cosmology, Department of Astronomy, University of Science and Technology \\of China, Hefei 230026, China\\
$^{2}$School of Astronomy and Space Sciences, University of Science and Technology of China, Hefei 230026, China\\
$^{3}$Kavli Institute for Astronomy and Astrophysics, Peking University, Beijing 100871, China\\
$^{4}$International Centre for Radio Astronomy Research, M468, The University of Western Australia, Crawley, WA 6009, Australia\\
$^{5}$Australian Research Council, Centre of Excellence for All Sky Astrophysics in 3 Dimensions (ASTRO 3D), Australia\\
$^{6}$National Astronomical Observatories, Chinese Academy of Sciences, 20A Datun Road, Chaoyang District, Beijing, China\\
$^{7}$Department of Astronomy, University of Massachusetts, Amherst, MA 01003, USA\\
$^{8}$Tsinghua Center of Astrophysics \& Department of Physics, Tsinghua University, Beijing 100084, China
}

\date{Accepted 2019 December 21. Received 2019 December 4; in original form 2019 June 12}

\pubyear{2019}

\begin{document}
\label{firstpage}
\pagerange{\pageref{firstpage}--\pageref{lastpage}}
\maketitle

\begin{abstract}
We study the relation between the properties of the bulge/disc components and the \hi mass fraction of galaxies. We find that at fixed stellar mass, disc colours are correlated with the \hi mass fraction, while bulge colours are not. The lack of a correlation between the bulge colour and the \hi mass fraction is regardless whether the bulges are pseudo, or whether the galaxies host bars or are interacting with a neighbour. There is no strong correlation between the colours of the discs and bulges either. These results suggest that the current total amount of \hi is closely related to the formation of discs, but does not necessarily fuel the formation of (pseudo) bulges in an efficient way. We do not find evidence for the star formation in the discs to be quenched by the bulges.
\end{abstract}




\section{Introduction}

Galaxy formation and evolution are complex processes that we do not yet totally understand. In the $\Lambda$ cold dark matter ($\Lambda$CDM) modeled universe, the first galaxies are formed by the collapsing and cooling of gas. After that galaxies assemble their mass and structure mostly via mergers in the early Universe. But the role of secular evolution becomes increasingly significant in the nearby universe, compared with the early Universe \citep{2004ARA&A..42..603K, 2007Natur.450.1020C, 2010ApJ...716..942F, 2010ApJ...723...54K}. Different generations of stars form and enrich the ISM with metals in secular processes.

The \hi is a key component in galaxy evolution, as it is the raw material from which molecular clouds and then stars form. In the past decade, several large single-dish \hi surveys have been carried out to understand how galaxies acquire the \hi gas, fuel their star formation, and evolve consequently.
The shallow blind surveys, HIPASS (\hi Parkes ALL-Sky Survey; \citealt{2001MNRAS.322..486B,2004MNRAS.350.1195M,2004AJ....128...16K}) and ALFALFA (Arecibo Legacy Fast ALFA; \citealt{2005AJ....130.2598G,2018ApJ...861...49H}), 
and the targeted, optically selected, GASS survey (GALEX Arecibo SDSS Survey, \citealt{2010MNRAS.403..683C,2018MNRAS.476..875C}) have provided a detailed quantification of \hi scaling relations \citep{1994ARA&A..32..115R}.

\cite{2010MNRAS.403..683C} showed that \hi gas fraction decreases strongly with stellar mass, stellar surface mass density and colour, but is only weakly correlated with galaxy concentration.
Both stellar mass surface densities and concentrations are correlated with the bulge-to-disc ratio of galaxies, therefore a direct link between the \hi content and the different structural components of the stellar part remains unclear.

It is also debated whether the link between the presence of bulges and galaxy quenching is casual or causal.
\citet{2009ApJ...707..250M} proposed that the bulge can regulate the efficiency of gas to form stars by increasing the shear between rotational orbits, and decreasing the mid-plane pressure of the stars. If this scenario works, then one would find an excess of \hi gas with respect to the star formation rate in galaxies which host prominent bulges.
However, \citet{2011MNRAS.411..993F} stacked the ALFALFA \hi spectra of early type galaxies, and find that early type galaxies show slightly lower $\fhi$ than galaxies which have similar colours and stellar masses. 
Their result does not strictly rule out the model of \citet{2009ApJ...707..250M}, because the bulges are not decomposed from the whole galaxy and  the properties of the bulges are not well quantified. 

The bulge-disc decomposition provides a useful way to parametrize the properties of the bulge. These parametrizations also bring new insights into galaxy evolution. For example,  \citet{2006MNRAS.371....2A} use GIM2D with a bulge plus disc fit 10095 galaxies from Millennium Galaxy Catalogue (MGC) and find that the galaxy colour bimodality is not due to two galaxy populations, but the two different component of one galalxy (i.e. red bulge and blue disc).   
\citet{2011ApJS..196...11S} increased the sample of \citet{2006MNRAS.371....2A} to more than 1 million galaxies and provide an extensive comparison set for detailed theoretical and observational studies of galaxies.

Bulges can also be broadly classified into two types: classical bulges and pseudo bulges. 
Classical bulges generally exhibit properties resembling elliptical galaxies,
which are more spherically symmetric, show a smooth distribution of stars, are supported mostly by velocity dispersion and have older stellar populations. 
 On the contrary, pseudo bulges show disk-like properties,  
have active star formation, are rotationally supported systems and show similar relations between the effective radius, the effective surface brightness and the central velocity dispersions as discs \citep{1993IAUS..153..209K,1994MNRAS.267..283A,2016ASSL..418...41F}.
 Classical bulges  follow a tighter relation between Sersic index and bulge-to-total ratio than that of pseudo bulges, and their mass-size relation is different from that of pseudo bulges which is closer to that of bars \citep{2009MNRAS.393.1531G}.
These differences in properties suggest that classical bulge are more likely to be the result of fast and violent processes, such as galaxy mergers, whereas pseudo bulges are more likely formed through slow, non-violent processes, like disc instabilities \citep{2004ARA&A..42..603K,2005MNRAS.358.1477A}.

As there are large differences between these two types of bulges, it is important to separate them in studies of the relation between bulges and other galactic properties. One convenient way is to allow the Sersic $n$ to vary (in contrast to fixing the Sersic $n$ to 4, which is equivalent to fitting a de Vaucouleurs classical bulge) in the bulge-disc decomposition procedure, and classify bulges with small $n$ ($<$2.5) as pseudo bulges and the rest as classical bulges \citep{2004ARA&A..42..603K}.

In this work we use the catalogue provided by \citet[][M15 hereafter]{2015MNRAS.446.3943M, 2016MNRAS.455.2440M}, which decomposed  $\sim$ 700,000 SDSS galaxies into disc and Sersic bulges, in the $r$, $g$ and $i$ bands. 
We study how the \hi gas fraction varies with bulge and disc propeties to explore the origin of galaxy bulges.

The paper is structured as follows. In Section 2 we describe the samples considered in this paper and their properties. 
In Section 3 we first study how the bulge colour varies with \hi fraction, then we examine how the structural parameters depend on \hi fraction, we perform a similar analysis for the discs, and also investigate the possible link between the colours of the discs and bulges.
Our conclusions and discussion are presented in Section 4.

\section{Sample and Data} \label{sec:sample}
\subsection{ALFALFA }
The ALFALFA survey is a blind 21 cm \hi survey that was carried out at the Arecibo 305 m radio observatory in Puerto Rico. It covers a large area of the sky at high Galactic latitude ($\sim$7000 deg$^2$), detecting \hi masses down to 10$^6$ $\msun$, with a velocity resolution of 11 $\rm{km~s^{-1}}$, and an angular resolution of $\sim$4\arcmin. The ALFALFA survey method, sensitivity limits, and potential uses of the data can be found in the \citet{2005AJ....130.2598G}.

ALFALFA  assigns a code to each galaxy indicating the confidence level of the detection. Code 1 objects have signal-to-noise ratio (S/N) values above 6.5; code 2 objects have 4.5 $<$ S/N $<$6.5, but possess a prior confirming the redshift. For our sample, we include galaxies (in total 31502) which have either code 1 or 2 from the 100\% catalogue of ALFALFA \citep{2018ApJ...861...49H} in our study.  We also test using code 1 only, but the main trends don't change significantly. We refer to them the ALFALFA  catalogue.

\subsection{The bulge-disc decomposition catalogue}
\citetalias{2015MNRAS.446.3943M} presents a catalogue of 2-dimensional, point spread function (PSF) corrected bulge-disc decomposition on the $r$-band images for  $\sim$ 700,000 spectroscopically selected galaxies drawn from the Sloan Digital Sky Survey (SDSS) Data Release 7.
They use a variety of models, including de Vaucouleurs, Sersic, de Vaucouleurs+Exponential, and Sersic+Exponential, in order to properly fit different types of galaxies.
For example, an elliptical galaxy is better described by a single Sersic component, a pure disk by an exponential component, while most of the others by Sersic+Exponential component.
Later they extend this catalogue to include the $g$ and $i$ bands \citep{2016MNRAS.455.2440M}.  They fit all bands separately, in contrast to using one band as the reference band which determines the structural parameters in the other bands \citep{2011ApJS..196...11S, 2012MNRAS.421.2277L, 2013MNRAS.430..330H}. By doing so, the decomposition can reflect the structural differences between stars of different ages, and the colour of different components can be more reliably derived. 

Here, we mainly use the sub-sample of galaxies which are best described by the Sersic+Exponential  model, as one goal of this paper is to study the relation between the different types of bulges and the \hi gas. 

\citetalias{2015MNRAS.446.3943M}  provides a series of flags to record the reliability of the decomposition for each galaxy. As we want to study the $g-r$ colour of bulge and disc separately, we only choose sources with flag 10 (good total and component magnitudes and sizes). 
We remove the galaxies with bulge effective radius $R_{\rm e,bulge}< 0.7\, \mathrm{arcsec}$, to ensure that the bulge can be resolved by the SDSS images which have a point spread function with a full width half maximum (FWHM) of ~1.4 arcsec.

Because \citetalias{2015MNRAS.446.3943M} fits the images of different bands seperately, we remove the galaxies which deviate more than $2\sigma$ from the mean relation of $\log(B/T)_{g}$  versus $\log(B/T)_{r}$.
We have 110,159 galaxies after all these selections. We refer to them the \citetalias{2015MNRAS.446.3943M}  bulge+disc (BD) sample.

It should be noted that the \citetalias{2015MNRAS.446.3943M} catalogue was produced with a pipeline for general galaxies, the decomposition may not be optimised for all the   galaxies.  This paper hence focuses only on a statistical analysis of the average trends of disc and bulge properties, without going into the details of individual systems.  \citet{2019MNRAS.490.4060C} investigated the influence of bulge-to-total mass ratio on the HI-richness of galaxies and did not find a significant effect, while in this paper we focus on the relation between the HI-richness of galaxies, and the growth ($g-r$ colour) of different types of bulges and disks.

\subsection{Working samples}

We start from the \citetalias{2015MNRAS.446.3943M}  BD sample (Section 2.2), select all galaxies with redshift $0.005<z<0.05$, stellar mass $M_{\star}>10^{9.5} M_{\sun}$, axis ratio $b/a>0.5$. 
This selection results in 9,543 galaxies, which makes the parent sample of this study.

 918 of the galaxies from the parent sample can be matched to the ALFALFA  catalogue (Section 2.1), with a maximum matching distance of 6 arcsec. 
We refer to them as Sample A. 

The Arecibo beam is $3.3 \times 3.8$ arcminutes, which is relatively large, as a result for each source of sample A, more than one galaxy may contribute to the 21 cm flux, leading to overestimate the \hi fluxes. 
To minimize the influence of contamination, we remove galaxies whose \hi flux may be contaminated by neighbouring galaxies within the beam from sample A following the method outlined in  \citet{2017MNRAS.464.3796T}. 
There are  60 galaxies which have blue companions (defined as $g-r<-0.01 \times (M_r+21)+0.65$) within one beam radius (1.9 arcminutes) and within a relative velocity of 250 $\mathrm{km~s^{-1}}$. These galaxies have high risk of being contaminated in their HI fluxes, and are removed. 
The remaining  858 galaxies are referred to as sample B, which is our main study sample in the following section.

 487 of the galaxies in sample B have metallicity measurements from the MPA/JHU catalogue \citep{2004ApJ...613..898T}, which we refer to as sample Metal.
We estimate the dust attenuation parameter E(B-V) for 796 galaxies from sample B, which have reliable $H\alpha$ and $H\beta$ flux measurements (with signal-to-noise ratio$>$3, from MPA/JHU). We refer to these  776 galaxies as sample EBV.
 631 of the galaxies in sample B can be found in Galaxy Zoo 2 \citep{2013MNRAS.435.2835W,2016MNRAS.461.3663H} and have a reliable identification for hosting bars or not. We refer to them as sample Bar. These samples are summarized in Table~\ref{table:1}. 

\begin{table}
\centering
\caption{Galaxy samples used in this work }
\label{table:1}
\begin{tabular}{lcl}
\hline
\hline
Sample &  N & Description \\
\hline
M15disk	&110159	&flag10 in \citetalias{2015MNRAS.446.3943M} , $R_{\rm e,bulge}<0.7''$,  outliers removed\\
Parent	&9543	&$M_\star>9.5 M_{\sun}$, $0.005<z<0.05$, $b/a>0.5$\\
A		&918		&Parent + \hi\\
B	&858		&A + remove \hi confused\\
Metal		&487		&gas phase metalicity in MPA/JHU\\
EBV		&776		&dust attenuation estimated from  $H\alpha / H\beta$\\
Bar		&631		& bar identification from Galaxy Zoo 2 \\
\hline
\end{tabular}
\end{table}

\subsection{Quantification of galaxy properties}
\begin{figure*}
	\includegraphics[width=0.95\linewidth]{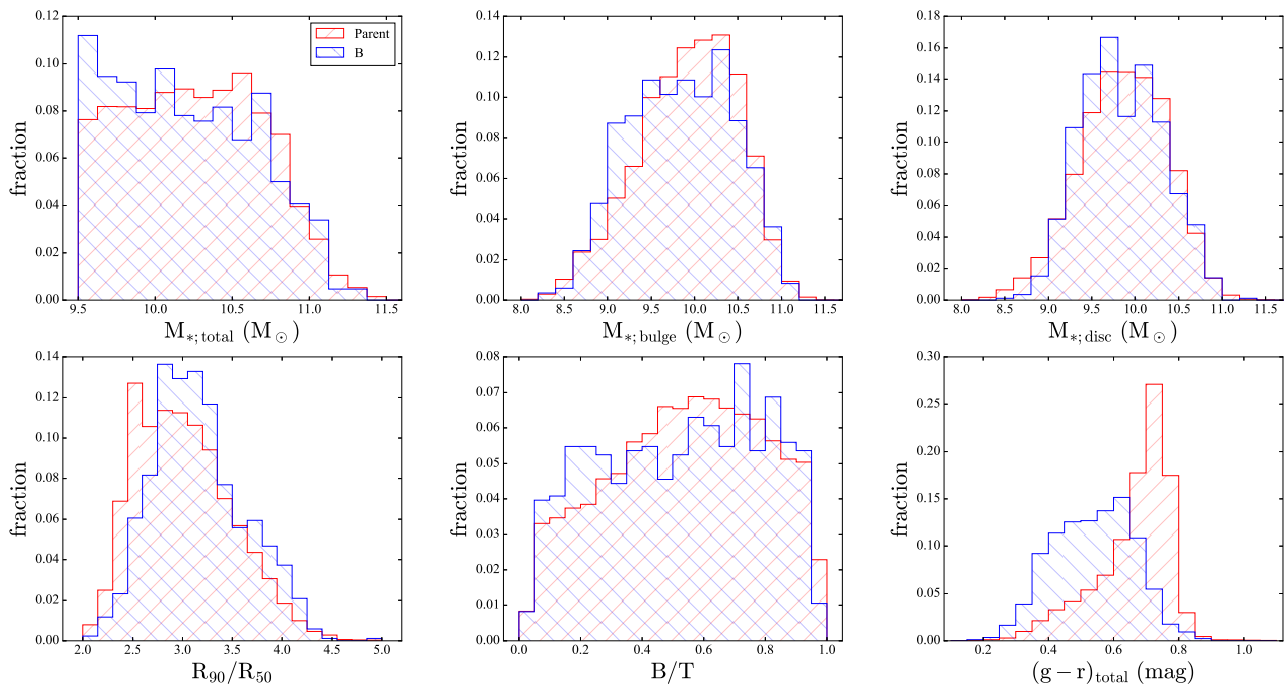}
    \caption{ Normalized distributions of stellar properties for the parent sample (red) and sample B (blue).}
    \label{fig:dis}
\end{figure*}

Here we introduce the parameters that we use to quantify galaxy properties, we also show the distribution of some properties for Sample B and the parent sample (see Fig.~\ref{fig:dis}). 
\begin{enumerate}
\item \hi mass: \hi mass is  calculated as ${\mhi=2.536\times 10^5~D_{\rm Mpc}^2 ~S_{21}} \msun$, where $D_{\rm Mpc}$ is the luminosity distance in unit of Mpc, and $S_{21}$ is the flux of the rest-frame 21 cm line in a unit of $\mathrm{Jy\,km\ s^{-1}}$ from the ALFALFA catalogue.

\item Colour: The total, bulge and disk $g-r$ colours, which are taken from  the \citetalias{2015MNRAS.446.3943M}  catalogue.

\item Stellar mass: The total, bulge and disk stellar masses, which are calculated via $\log_{10}(M/L)=a_{\rm \lambda}+b_{\rm \lambda} \times (g-r)$ \citep{2009MNRAS.400.1181Z}. We use the $r$-band luminosity, $a_\lambda=-0.840$ and $b_\lambda=1.654$. 
Because we calculate the colour and mass for the bulge and disc separately, the sum of masses in the bulge and in the disc may differ (by on average 0.10 dex, with a scatter of 0.11 dex) from the mass estimated based on the luminosity of  the whole galaxy. For consistency with the total stellar mass, we normalize the bulge mass as 
$\mb = \mt \times \mb /(\mb+\md)$. We perform similar normalization for the disc stellar mass.

\item Stellar surface mass density: this quantity is defined as $\mu_\star=M_\star/(2\pi R_{\rm 50,r}^2)$, where $R_{\rm 50,r}$ is the $r$ band half-light radius.  We calculate $\mu_\star$ for the discs and bulges, using the $R_{\rm 50,r}$ of each component separately. 

\item Concentration: defined as $R_{\rm 90,r}/R_{\rm 50,r}$, where $R_{\rm 50,r}$ and $R_{\rm 90,r}$ are radii containing 50 and 90 per cent of the Petrosian flux in $r$ band respectively.

\item $B/T$: the bulge-to-total mass ratio, which describes the bulge fraction of galaxies. 
\item Bulge Sersic $n$ index: the Sersic model index $n$ in the r-band. Bulges which have Sersic $n<2.5$ are classified as pseudo bulges and the rest classical bulges \citep{2004ARA&A..42..603K}.

\item Gas phase metallicity: from the MPA/JHU catalogue, where the oxygen abundance was estimated using the models in \citet{2001MNRAS.323..887C} as discussed in  \citet{2004ApJ...613..898T}.

\item Bar properties: from the Galaxy Zoo 2 sample. We select the barred galaxies which have a probability of being barred greater than 0.5, following the criteria of \citet{2011MNRAS.411.2026M}. There are 238 barred galaxies in Sample A.
\item Galaxies experiencing galactic interactions: We use two parameters to indicate possible galactic interactions for each galaxy. The first is the  distance to the nearest neighbour galaxy $D_{\rm neighbour}$, and the second is the nonparametric morphological  parameters G (the Gini coefficient)  and A (the morphological asymmetry) \citep{2010MNRAS.409.1379V}. A small value of $D_{\rm neighbour}$ indicates an early stage of possible interactions.
A large value of G indicates the flux to be concentrated in a small fraction of  pixels, while a large value of A indicates significant asymmetry. A combination of the two parameters makes a sensitive diagnostic for on-going interactions \citep{2004AJ....128..163L}. 
We use $D_{\rm neighbour}<0.25$  \citep{2008MNRAS.385.1903L} and $G>-0.4\times A+0.66$ or $A\geq 0.4$ \citep{2004AJ....128..163L} as two independent criteria to indicate interactions. We note that the number of galaxies meeting the second criteria is small because of the selection bias toward galaxies which have well decomposed bulge and disc components. 

\end{enumerate}

We can see from Fig.~\ref{fig:dis} that sample B is biased toward bluer galaxies, due to the selection of galaxies with ALFALFA detections. The analysis in this paper hence only applies to relatively star-forming galaxies. The sample also misses the galaxies which have the lowest $R_{\rm 90,r}/R_{\rm 50,r}$, because we require the presence of bulges.

\subsection{Correlation analysis method}\label{sec:method}
We use correlation coefficients to study the link between properties. We use the Pearson correlation coefficient to indicate the strength of linear correlation between two parameters.  Positive and negative coefficients indicate positive and anti-correlations, respectively. The absolute value ranges from 0 to 1, and absolute values above 0.4 are considered to indicate a relatively strong correlation compared to the relation between randomly distributed parameter pairs. 
We also calculate the Spearman correlation coefficient as supplementary information to the Pearson correlation coefficient. It quantifies the correlation between the ranks of two parameters, so the correlation is not necessarily linear. 

Because many galactic properties correlate with each other (e.g. both colour and $\fhi$ are correlated with $M_\star$ in star-forming galaxies), there is possibility that a correlation between two parameters  is  caused by dependence of both parameters on a third parameter. Here we rely on two ways to identify a possible intrinsic (independent of parameter $C$) linear correlation between two parameters $A$ and $B$.
In the first way, we calculate the partial correlation coefficient. The partial correlation coefficient between A and B, with a third control parameter C, is defined as $\rho_{A B, C}=\left(\rho_{A B}-\rho_{A C} \rho_{B C}\right) / \sqrt{\left(1-\rho_{A C}^{2}\right)\left(1-\rho_{B C}^{2}\right)}$, where $\rho_{A B}$ is the Pearson correlation correlation between $A$ and $B$ (similar for  $\rho_{A C}$ and $\rho_{B C}$). 
We use two methods to justify the significance of correlations. Firstly, we calculate errors of the coefficients through bootstrapping. We resample the data with replacement for 2000 times, and thereby build 2000 new samples. We calculate the correlation coefficient for each new sample and thereby obtain a distribution of 2000 correlation coefficients. The standard deviation of the distribution is taken as the error of the correlation coefficient of the original sample.   Secondly, we use the python package \verb#Pingouin# \citep{Vallat2018} to calculate a 95\% parametric confidence interval ($\pm$2 sigma) and two-tailed non-hyopthesis p-value for each correlation coefficient. Higher absolute values of correlation coefficients (after considering the uncertainties quantified by error bars or confidence intervals) should indicate stronger correlation. We also find that all the significant correlations ($\rho>0.4$) in this study have a p-value less than 0.001, indicating that the null hypothesis of no correlations is rejected with a 0.001 significance level ($>3$ sigma confidence level), and hence these correlations are indeed significant. Some of the very weak correlations, instead, have high p-values (e.g. the partial correlation between $\fhit$ and $\grd$ controlled by $\mt$, see Section \ref{sec:result}). We don't consider  parameter uncertainties during the calculation of correlation coefficient, but in sample selection we have removed data with large error bars.

In the second way, we divide the sample into sub-samples by the control parameter $C$, and plot the median relation between $A$ and $B$ in each sub-sample. If the correlation between $A$ and $B$ is independent of $C$, we would expect the median relations of different $C$ bins to be close to each other. On the contrary, if the $A$ versus $B$ relation of different $C$ bins differ significantly from each other and even have no overlapping between the adjacent $C$ bins, then it would indicate the intrinsic correlation between $A$ and $B$ to be very weak (even if they appear to be correlated in each individual $C$ bin).

\section{Results} \label{sec:result}
To explore the relation between the \hi gas and the formation of discs and bulges, we study the dependence of the colours of the discs and the bulges on the \hi gas fraction. \hi is the reservoir of material for forming stars, and colour indicates the specific star formation rate (defined as $\mathrm{SFR}/M_\star$ the star formation rate over the mass of stars formed). A strong correlation between the colour of a stellar component and the \hi gas fraction will indicate a strong link between the formation of this stellar component and the \hi gas. 

\begin{figure*}
	\includegraphics[width=0.95\linewidth]{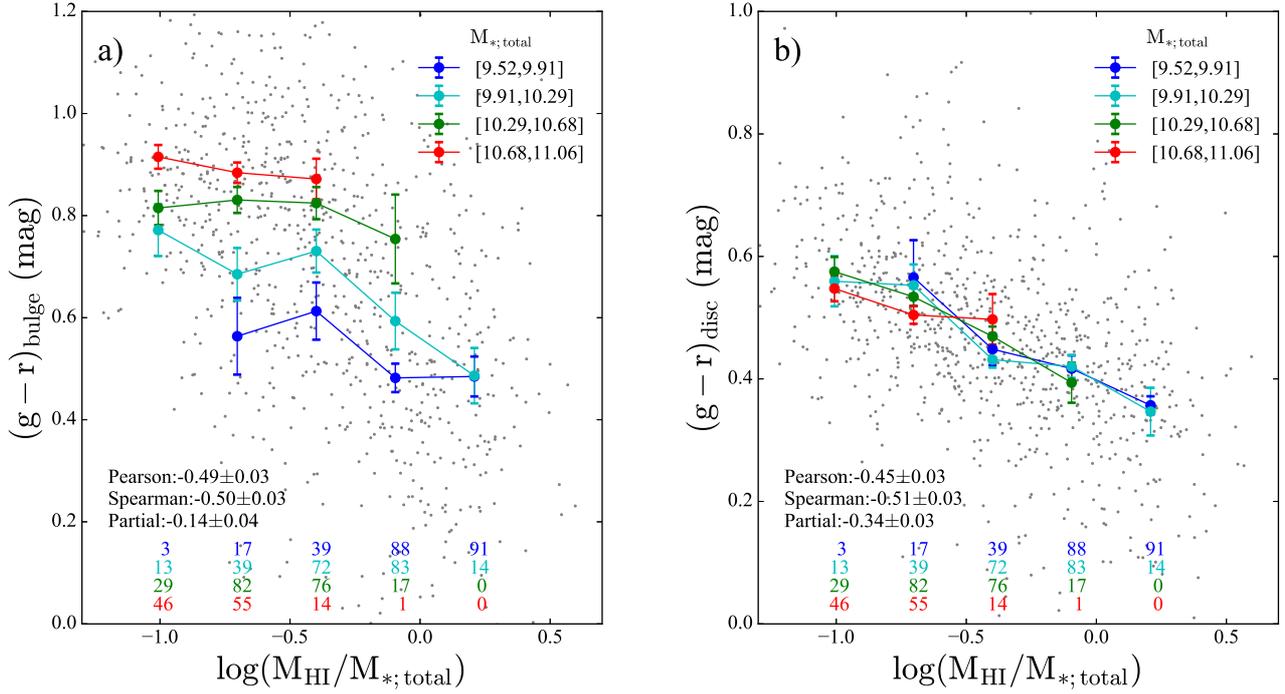}
    \caption{The dependence of bulge (left) and disc (right) colours on $\fhi$. The sample is divided into sub-samples by the stellar mass, as denoted in the top-right corner of each panel.  The large solid dots and error bars are the the median value and its error derived through bootstrapping. These coloured dots provide a visual impression of the strength of correlations (see Section 2.5). The number of galaxies in each sub-sample is denoted in the bottom of each panel.  Pearson, Spearman and partial correlation (with the effect of the stellar mass removed) coefficients of all the data points are shown on the  bottom  left of each panel.}
    \label{fig:color_total}
\end{figure*}

In Fig.~\ref{fig:color_total}, we show how the  bulge and disc colour as a function of $\fhi$. The sample is divided into sub-samples by stellar mass, as denoted in the top-right corner of each panel, different colours indicating different stellar mass bins. For each $M_\star$ divided sub-sample, we further divide it into different $\fhi$ bins. We exclude bins with less than 10 galaxies to ensure statistical significance. We use the median value and its error derived through bootstrapping (large solid dots and error bars) to quantify the distribution of data points in the bin. The trend of these median values give a visual impression how strongly parameters are correlated (supplementary to the quantification with partial correlation coefficients, see Section~\ref{sec:method}). Replacing median values with mean values does not significantly change the overall trend. We also show the Pearson, Spearman and partial correlation coefficients in the bottom-left corner of each panel. These correlation coefficients for all the parameter pairs investigated in this paper, as well as their 1-$\sigma$ error bars, 95\% confidence intervals, and p-values, are summarized in  Table~\ref{table:2}.

At first glance, there is a considerable anti-correlation between the bulge colour and $\fhi$, with a Pearson correlation coefficient of $-0.49$. 
However, the bulge colours are nearly constant as a function of the $\fhi$ in each stellar mass bin, and we only get a partial correlation coefficient of $-0.14$ between them after removing the effect of the stellar mass.
This means that when the stellar masses are controlled, there is no intrinsic correlation between bulge colour and $\fhi$. 

On the other hand, for the disc component, there is considerable correlation between $\fhi$ and colour in each stellar mass bin. The partial coefficient is $-0.34$ with the effect of the stellar mass removed. So dependencies on $M_\star$ could not fully explain the link between the disc colour and $\fhi$.

\begin{figure*}
	\includegraphics[width=0.95\linewidth]{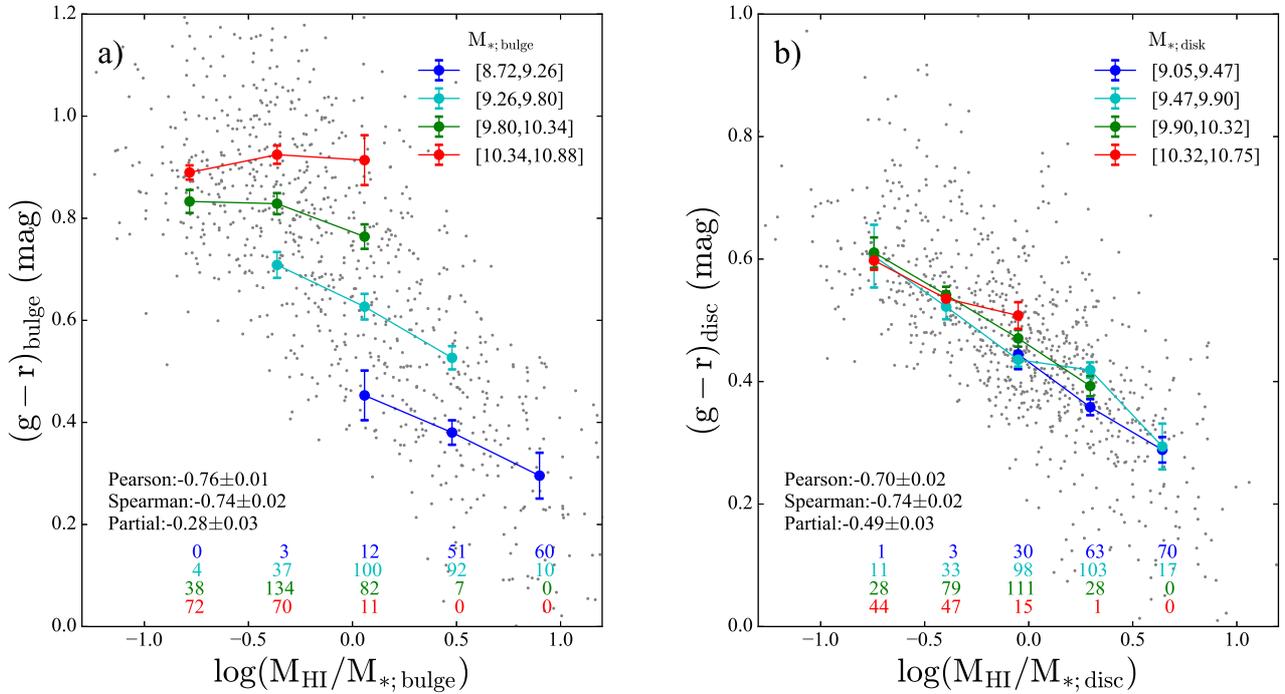}
    \caption{The dependence of bulge colours on $\fhib$ (left) and disc colours on $\fhid$ (right) (see Sect. 3). Similar as Fig.~\ref{fig:color_total}, except that $\fhi$ is replaced by $\fhid$ or $\fhib$.}
    \label{fig:color_component}
\end{figure*}

We find that $\fhi$ may be more closely related to the stellar population in the disc than that in the bulge.
We perform the following tests to further check this result. 
We assume that all \hi is associated only with the disc but not with the bulge, and define a new parameter $\fhid=\mhi/\md$. We check this assumption in  Fig.~\ref{fig:color_component}, which is similar as Fig.~\ref{fig:color_total}, except that $\fhi$ is replaced by $\fhid$ or $\fhib$. If this assumption is correct, we would find a tighter correlation of the disc colour versus $\fhid$, than versus $\fhi$. This is indeed what we find in Fig.~\ref{fig:color_component}(b).
Moreover when removing the effect of the disc stellar mass, the correlation is still significant with partial correlation of $-0.49$.
We do a similar test with the bulge, and define a new parameter $\fhib=\mhi/\mb$. As shown in Fig.~\ref{fig:color_component}(a), the partial correlation between $f_{\hi,bulge}$ and bulge colour is as weak as the partial correlation between $\fhi$ and the bulge colour.


Hence we find that for both discs and bulges, the colours are correlated with the stellar masses, but the former also show correlation with $\fhi$ while the latter does not.
It is not surprising to find such different dependences of bulges and discs on $\fhi$, as bulges can be classical and be formed via mergers of other galaxies \citep{1993MNRAS.264..201K, 1996MNRAS.283.1361B}, 
or the infall of massive star-forming clumps at high redshift \citep{2008ApJ...687...59G,2014ApJ...780...57B,2013MNRAS.432..455D,2013MNRAS.436..259P},
and then have no direct relation with the current status of the \hi gas. In the following sections, we specifically select the bulges that are likely to be secularly formed and investigate their relation with the \hi gas. We also investigate how the correlation between disc colour and $\fhi$ is affected by the structural properties of the disc and the bulge.


\subsection{The bulge colour and the HI gas} 
\begin{figure*}
	\includegraphics[width=0.95\linewidth]{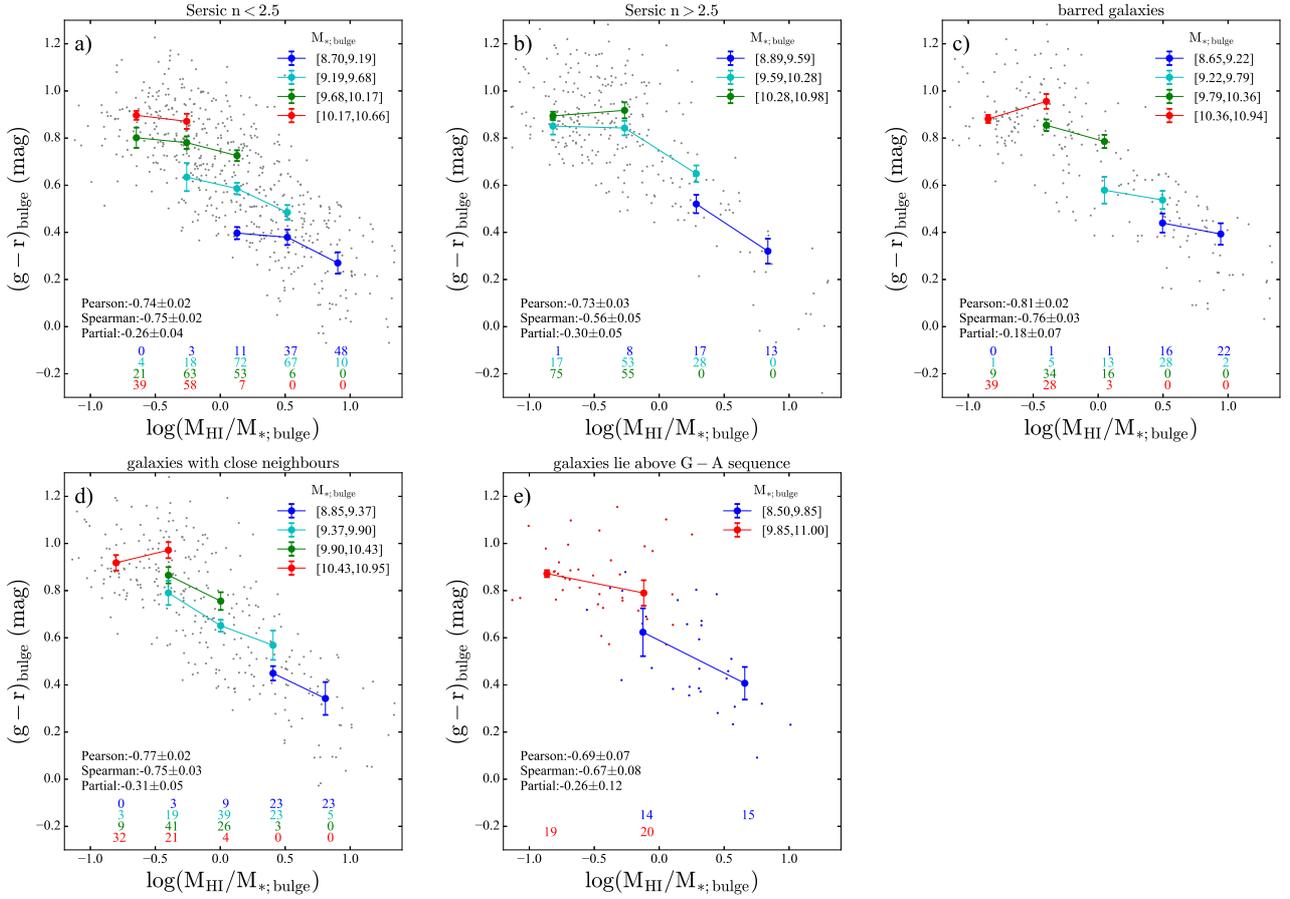}
    \caption{The dependence of  bulge colours on $\fhi$. Similar as the left panel of Fig.~\ref{fig:color_total}, but the sample is replaced by galaxies with pseudo bulges (a), galaxies with classical bulges (b), galaxies with bars (c) and galaxies experiencing galactic interactions (d,e). 
The sample is divided into sub-samples by the bulge mass, as denoted in the top-right corner of each panel.  Same symbols and errorbars as Fig.~\ref{fig:color_total}.}
    \label{fig:diff_bulge}
\end{figure*}

We have shown that the bulge colour depends more strongly on $\mb$ than on $\fhi$. We investigate whether this result holds for the secularly evolving pseudo bulges in this section. We hence plot  the dependence of bulge colours on $\fhi$ for different types of galaxies, as shown in Fig.~\ref{fig:diff_bulge}.

Pseudo bulges are disk-like while classical bulges are not; including the classical bulges may have smeared a possible correlation between the colour of the pseudo bulge and $\fhi$. So we divide the sample by bulge type, where the pseudo bulges have Sersic $n<2.5$ and the others are classical bulges (see Section 2).
However, as shown in Fig.~\ref{fig:diff_bulge}(a) and (b), the bulge colours depends more strongly on $M_\star$ than on $\fhib$, for the mean $\grb-\fhib$ relations of adjacent $M_\star$ bins depart significantly from each other in the intercepts, e.g. the reddest $\grb$ of the blue curve is bluer than the bluest $\grb$ of the cyan curve.

Bars are an efficient way of driving gas inflows and secularly forming the central bulges \citep{2012MNRAS.423.3486W}, so we would expect that a correlation between the stellar population and $\fhi$ may be found for the barred galaxies. But we don't see any significant trend between the bulge colours and $\fhi$ at fixed bulge stellar mass for barred galaxies (Fig.~\ref{fig:diff_bulge} (c)). 

Tidal interactions are also an efficient way to drive gas inflows and enhance the central star formation \citep{1972ApJ...178..623T,1991IAUS..146..243K,1999PhR...321....1S}. As described in Section 2.4, we use neighbour counts and morphological parameters to select the potentially interacting galaxies. Both parameters are found in the literature to be correlated with enhanced central SFR \citep{2008MNRAS.385.1903L,2009ApJ...691.1005R}. When the samples are small,  we show trends with fewer $\mb$ and $\fhib$ bins. We find no significant correlation between the bulge colour and $\fhi$ for the interacting galaxies (Fig.~\ref{fig:diff_bulge} (d,e)).

\begin{figure*}
	\includegraphics[width=0.95\linewidth]{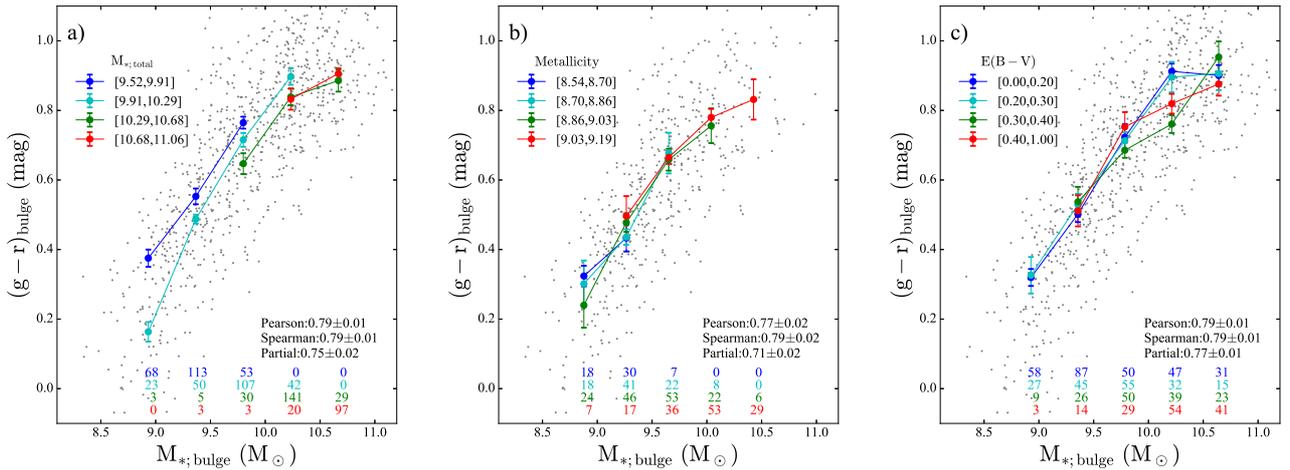}
    \caption{The dependence of $\grb$ on the bulge mass. The sample is divided into sub-samples by the bulge mass, gas phase metallicity, or E(B-V), as denoted in the top-left corner of each panel.  Same symbols and errorbars as Fig.~\ref{fig:color_total}.}
    \label{fig:bulge_color}
\end{figure*}

One question is whether the bulge colour could be more closely related to $\mt$ than to $\mb$, as galaxy evolution strongly depends on the total stellar mass \citep{2003MNRAS.341...54K}, and massive galaxies tend to have more massive and redder bulges \citep{1996AJ....111.2238P, 2000MNRAS.312..497B}.
In Fig.~\ref{fig:bulge_color} (a), we show that the bulge colour is more strongly correlated with the bulge mass than the total stellar mass.
It was hence proper to use $\mb$ as a control parameter when investigating the relation between the colour and $\fhi$ of bulges.
 
Another possible worry is that the bulge colours used here may actually reflect the attenuation or gas phase metallicity, as colours not only depend on the stellar age, but also on these two properties \citep{1999PASP..111...63F,1994ApJS...95..107W}. In Fig.~\ref{fig:bulge_color} (b) and (c), we show that neither the metallicities nor the dust attenuations can fully explain the correlation between the bulge colours and the bulge masses. It supports the usage of colour as an indicator of the star-forming status in the bulges.

In summary, we find that for all the different sub-samples, the star formation in the (pseudo) bulges is strongly correlated with the stellar mass of the bulges, but not correlated with $\fhi$ of the galaxies. We note that if we replace $\fhi$ by $\fhib$, the general trends in this section do not significantly change.

\subsection{The disc colour and the HI gas}

We have shown that $\fhi$ is not correlated with the bulge colour, but strongly correlated with the disc colour.  This suggests that the \hi content is more closely related to the disc component than the bulge component. We therefore use $\fhid$ to investigate how the formation of the disc is affected by structural properties of the disc and the bulge. 

It is unclear whether  bulges directly play a role in the evolution of galaxies by stabilizing the discs with their centrally concentrated potential \citep{2009ApJ...707..250M}. So we investigate whether properties of the bulge have an influence on the relation between the disc colour and $\fhid$. If the bulge quenching mechanism plays a major role in reducing the star forming efficiency in our sample, we would expect to observe an on average redder $\grd$ at a fixed $\fhid$ for galaxies with significant bulges than for galaxies with less significant ones.

\begin{figure*}
	\includegraphics[width=0.95\linewidth]{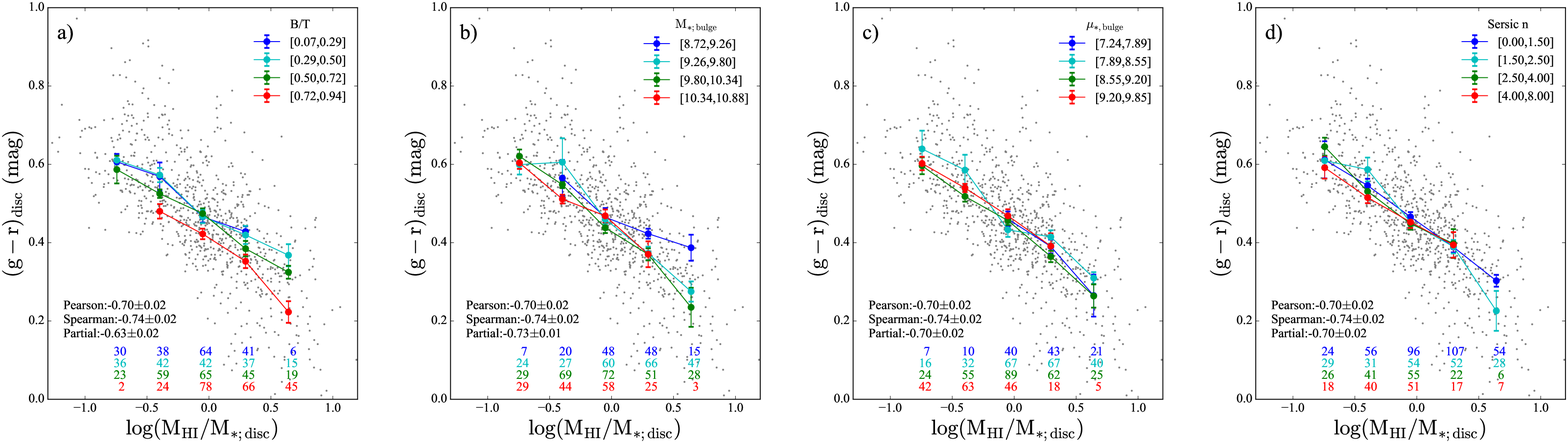}
    \caption{The dependence of disc colour on $\fhid$. The sample is divided into sub-samples by the $B/T$ (a), stellar masses (b), stellar surface mass densities (c) of bulges and  bulge Sersic n index (d).  Same symbols and errorbars as Fig.~\ref{fig:color_total}.}
    \label{fig:bq_1}
\end{figure*}

 From Fig.~\ref{fig:bq_1}, we can see that the partial correlation coefficients between $\grd$ and $\fhid$ are between -0.63 and -0.73 for different control parameters ($B/T$, $\mb$, $\mu_{\rm \star, bulge}$ or Sersic $n$), very close to the Pearson correlation coefficient of -0.70. It suggests that disc colours do not depend on $B/T$, $\mb$, $\mu_{\rm \star, bulge}$ or Sersic $n$. Therefore we do not find evidence for the disc colours to be affected by the existence or properties of bulges.




\subsection{The relation between the bulge colour and the disc colour}
\begin{figure*}
	\includegraphics[width=0.95\linewidth]{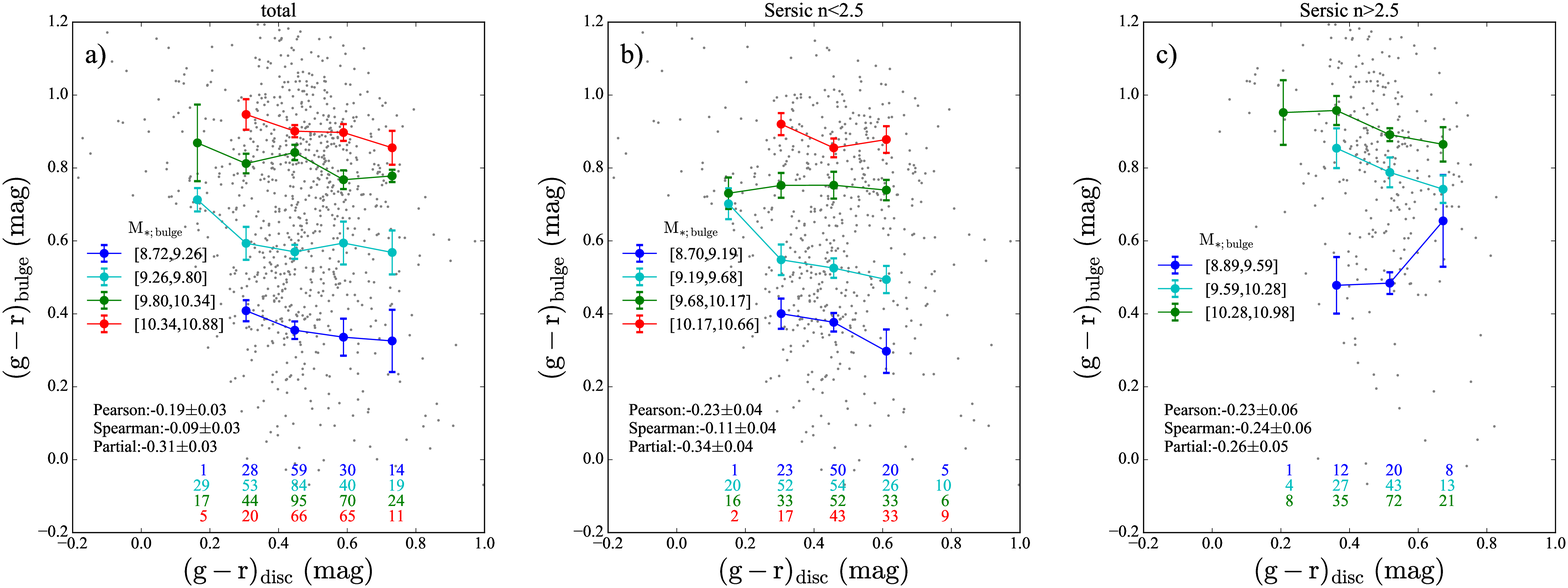}
    \caption{The dependence of bulge colour on disc colour for total, pseudo bulge ($n<2.5$) and classical bulges ($n>2.5$) respectively. Different colours indicate different bulge stellar mass bins, as denoted in the middle-left corner of each panel.  Same symbols and errorbars as Fig.~\ref{fig:color_total}.}
    \label{fig:disk_bulge_coevolutin}
\end{figure*}

In previous sections, we have shown that the bulge colour is largely independent of $\fhi$, while the disc colour is strongly correlated with $\fhi$. These results indicate that the bulge and the disc evolve in different ways, even when both are evolving (mostly) secularly at low redshift. 
We confirm this point in Fig.~\ref{fig:disk_bulge_coevolutin}, that  bulge colours are not correlated with  disc colours in a positive way, when the bulge mass is controlled.

\begin{figure*}
	\includegraphics[width=0.95\linewidth]{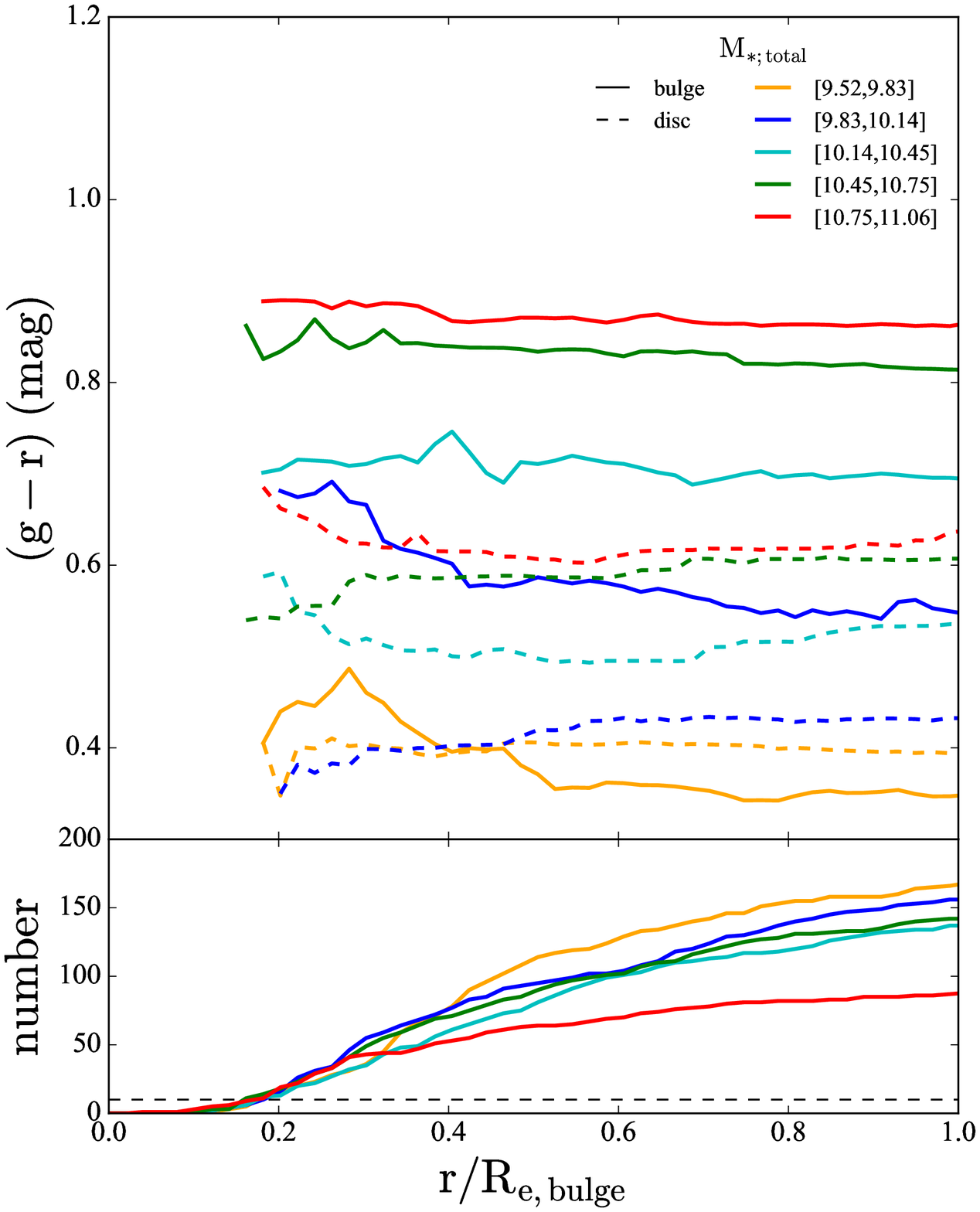}
    \caption{Top: The median colour profiles for the bulge (solid) and disc (dashed) components in different total mass bins. For each galaxy, only data points with radius larger than 0.5 times the FWHM of SDSS are used. Different colours indicate different stellar mass bins, as denoted in the top-right corner. The profile is plotted out to the effective radius of the bulge; Bottom: The  number of galaxies in each radius bin. The dashed horizontal line is at a level of 10, which is the minimum number of galaxies  to calculate median colour profile.}
    \label{fig:radial}
\end{figure*}

It is possible that the bulge and the disc differ in colours due to different characteristic sizes. So we compare their colour radial distributions. 
We plot median colour profiles for the bulge  and disc  components in each total mass bin.
In Fig.~\ref{fig:radial}, we can see that at a fixed radius, there is indeed a general trend for galaxies which have high $\mt$ to have redder colours in both the bulges and the discs than the galaxies with low $\mt$. However, at a fixed radius, bulge colours vary much more significantly as a function of $\mt$ than disc colours. When $r=0.5R_{\rm e}$, the average bulge colours increase from 0.4 to 0.9 mag when $\mt$ increases, but disc colours only increase from 0.4 to 0.65 mag. In a fixed $\mt$ bin, the average disc colours also vary much less significantly as a function of radius than bulge colours. All these results suggest a significant difference in the stellar population between the bulge and the disc.

\subsection{$\fhi$ and the bulge/disc structures}
 \citet{2010MNRAS.403..683C} showed that $\fhi$ is correlated with the stellar mass surface density but not  with the  concentration, hence we study how  bulge or disc structures affect $\fhi$ in a decomposed view.

\begin{figure*}
	\includegraphics[width=0.95\linewidth]{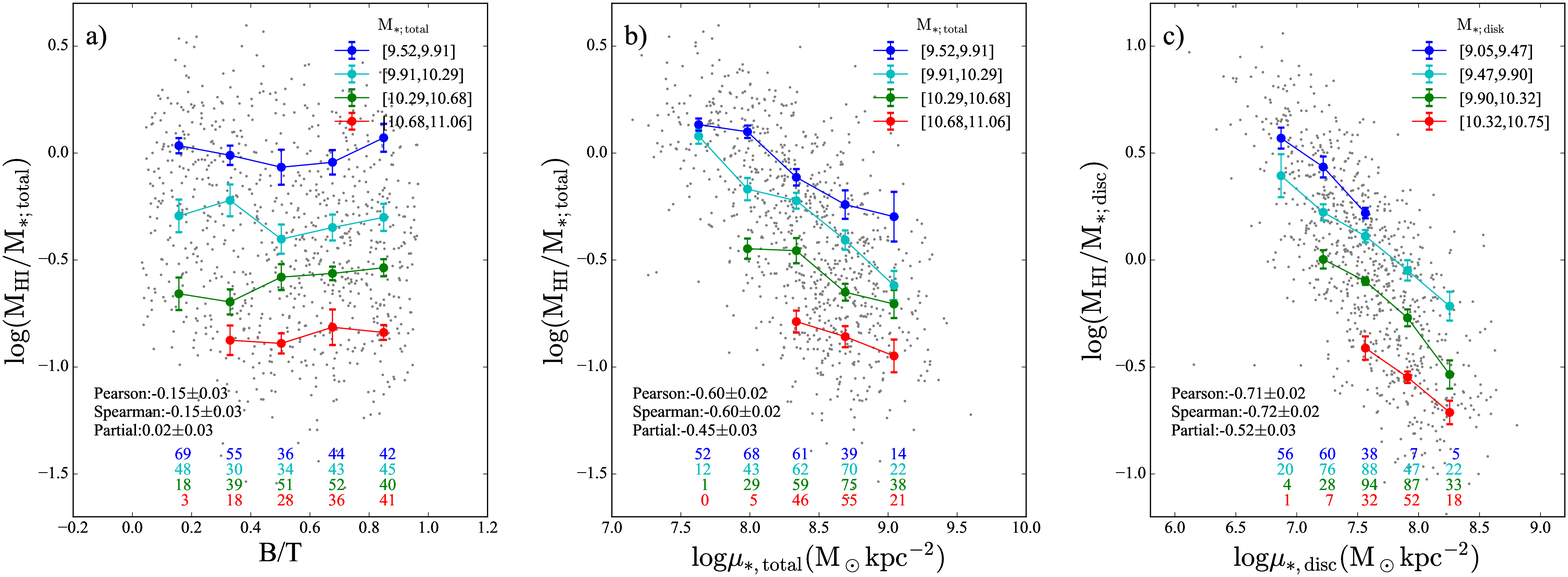}
    \caption{The dependence of $\fhit$ on $B/T$ (a), $\fhit$ on total surface mass density (b) and $\fhid$ on disk surface mass density (c). The sample is divided into sub-samples by the total stellar masses (a,b) or disc stellar masses (c). Different colours indicate different disc stellar mass bins.  Same symbols and errorbars as Fig.~\ref{fig:color_total}.}
    \label{fig:structure}
\end{figure*}

Firstly, from Fig.~\ref{fig:structure}(a), we shown that $\fhi$ is independent of $B/T$, which is consistent with the finding  that $\fhi$ only shows a weak correlation with concentration, as $B/T$ is tightly correlated with the  concentration. 
Hence the trend of decreasing $\fhi$  with increasing  $\mu_\star$ (\citealt{2010MNRAS.403..683C},  also panel b of Fig.~\ref{fig:structure}) is unlikely caused by the increased significance of bulges, although $\mu_\star$ is correlated with $B/T$. Instead, we also find a correlation between $\fhid$ and $\mu_{\rm \star, disc}$ (panel c of Fig.~\ref{fig:structure}), indicating that the disc structure  itself may affect or reflect the $\hi$ content of galaxies. 

\section{Summary and discussion}  \label{sec:discussion}
In this work, we used a sample of 858 SDSS galaxies with \hi data from ALFALFA and photometric bulge-to-disc decompositions from \citetalias{2015MNRAS.446.3943M}  to investigate  the relation between \hi content and bulge/disc properties.
We found that both the disc colour and the bulge colour are strongly correlated with $\fhi$, but for different reasons. The correlation between the disc colour and $\fhi$ is largely independent of the disk stellar mass. The correlation between the bulge colour and $\fhi$ is intrinsically weak, and is caused by both parameters depending on the stellar mass of the bulge. Consistently, the colours of the bulge and the disc are not significantly correlated when the bulge mass is controlled. These results about bulges hold when only pseudo bulges are considered. The results suggest the different ways that bulges and discs form.

In cosmological models, discs form inside-out as a result of gas accretion from the circum-galactic medium \citep{1996MNRAS.281..475K,1998MNRAS.295..319M}. The strong correlation between disk colours and $\fhi$ is consistent with this model. This model was supported by earlier observations that the colours of the outer regions of galaxies depend more strongly on $\fhi$ than the inner regions \citep{2011MNRAS.412.1081W}. In this paper, instead of dividing galaxies into the bulge and disc dominated inner and outer regions, we directly use the colours of disc components and find a strong correlation with $\fhi$, providing more direct support to the inside-out formation model than before. Moreover, we do not find evidence for the star formation in the discs to be quenched by the bulges.

It is well accepted that pseudo bulges are different from classical bulges but similar to discs. They tend to be star-forming and distribute similarly in the structural fundamental plane as discs \citep{2009ApJ...707..250M,2009MNRAS.393.1531G}. One may hence expect the star formation rate of pseudo bulges to be also correlated with the \hi content of galaxies. However, we find this correlation to be weak. 
The reason is likely that the \hi gas needs time to flow inward in order to reach the bulge. Effective inflows of gas can be driven by torques provided by bars (as a non-axisymmetric structures in galaxies), or external tidal effects \citep{2004ARA&A..42..603K}. The strength of tidal effects depend on the mass and orbit of the interacting companions \citep{2002MNRAS.332..155L}, and the strength of bars depend on the length, ellipticity and mass of bars \citep{2012MNRAS.423.3486W, 2004ApJ...600..595R}. The inflow rates of \hi hence may differ significantly among galaxies which have bar or are interacting, and the correlation between the bulge colour and $\fhi$ therefore may tend to have large scatter in these galaxies. Additional scatter between the colour and $\fhi$ could be caused by the fact that star formation triggered by strong inflows could be episodic, due to the high central mass concentration (hence strong shearing), quick consumption of gas, and$\slash$or strong stellar feedback \citep{2002MNRAS.329..502M, 2005ApJ...632..217S,2015MNRAS.453..739K}. As a result of these complexities, the fraction of the \hi that actually reaches the bulge region is uncertain, and the formation of the pseudo bulge can be temporarily detached from the formation of the discs. 

In semi-analytical models, the formation of a pseudo bulge is often empirically treated as the result of disc instabilities that emerge when a certain critical density is reached in the disc \citep[e.g.][]{2012MNRAS.422..997K}. The critical densities are reached when significant amount of gas flows inward, and hence is not directly related to the integrated gas fractions. This empirical treatment is qualitatively consistent with the way that pseudo bulges form as indicated by the results of this paper. But more details about the formation of pseudo bulges that can be compared with or constrain the cosmological models are largely missing. These details can be quantified as the inflow rate of gas to the bulge regions and the consequent surface density distribution of cold gas in the bulge regions; also relevant are the circular velocity and velocity dispersion of the cold gas that may affect the efficiency of converting the gas to stars of the bulges. These measurements need be extracted from high-resolution images of cold gas, for statistically significant samples of galaxies, which may be available in the coming years when large HI surveys with the Square Kilometre Array and its  precursor telescopes start.
 
We point out that our sample only includes relatively star-forming galaxies which have ALFALFA detections. The results relevant to the quenching of star formation rate (bulge quenching) could be affected by this selection effect. Also, we only investigated the relation between the HI and star formation rate (SFR), because we do not have data for the molecular gas.  It is possible that the conversion between HI and H2 or between H2 and SFR could be more affected by the bulge than the less direct relation between the HI and SFR. Because the bulge-disc decomposition catalogue used in this paper was produced by a pipeline \citep{2010MNRAS.409.1379V}, large uncertainties in the fitting could exist for individual galaxies. Our investigation is hence limited to the statistical aspect. Analysis based on deeper \hi data \citep[the x-GASS sample][]{2018MNRAS.476..875C}, as well as more careful component decomposition for individual galaxies, is currently under way \citep{2019MNRAS.490.4060C}.

\section*{Acknowledgements}
We thank the anonymous referee for constructive comments. We gratefully thank R. Cook, Y. H. Zhao, T. Xiao, Z. S. Lin for useful discussions.
Parts of this research were supported by the Australian Research Council Centre of Excellence for All Sky Astrophysics in 3 Dimensions (ASTRO 3D), through project number CE170100013. 
This work was supported by the National Science Foundation of China (NSFC, Nos. 11421303, 11433005, 11721303, 11973038, and 11991052) and the National Key R\&D Program of China (2016YFA0400702, 2017YFA0402600).
\begin{landscape}
\begin{table}
\centering
\caption{The Pearson, Spearman and Partial correlation coefficients  parameters investigated in this paper, as well as their 1-$\sigma$ error bars, 95\% confidence intervals, and p-values. }
\label{table:2}
\begin{tabular}{lllllccccccccc}
\hline
\hline
\multirow{2}{*}{Fig} &
\multirow{2}{*}{A (x axis)}&
\multirow{2}{*}{B (y axis)}&
\multirow{2}{*}{C (control parameter)}&
\multirow{2}{*}{Sample}&
\multicolumn{3}{|c|}{Pearson}&
\multicolumn{3}{|c|}{Spearman}&
\multicolumn{3}{|c|}{Partial}\\
&&&&      &r &CI &p-value  &r &CI &p-value &r &CI &p-value\\ 
\hline
2-a &$\mathrm{\log ({M_{HI}/M_{*,total}})}$ &$\mathrm{(g-r)_{bulge}}$ &$\mathrm{M_{*,total}}$ &B &-0.49$\pm$0.03 &[-0.54,-0.44] &7.19e-53 &-0.50$\pm$0.03 &[-0.55,-0.45] &5.49e-56 &-0.14$\pm$0.04 &[-0.20,-0.07] &6.41e-05\\
2-b &$\mathrm{\log ({M_{HI}/M_{*,total}})}$ &$\mathrm{(g-r)_{disc}}$ &$\mathrm{M_{*,total}}$ &B &-0.45$\pm$0.03 &[-0.51,-0.40] &8.89e-45 &-0.51$\pm$0.03 &[-0.56,-0.46] &7.00e-59 &-0.34$\pm$0.03 &[-0.40,-0.28] &9.68e-25\\
3-a &$\mathrm{\log ({M_{HI}/M_{*,bulge}})}$ &$\mathrm{(g-r)_{bulge}}$ &$\mathrm{M_{*,bulge}}$ &B &-0.76$\pm$0.01 &[-0.79,-0.73] &6.31e-165 &-0.74$\pm$0.02 &[-0.77,-0.71] &9.23e-152 &-0.28$\pm$0.03 &[-0.34,-0.21] &1.65e-16\\
3-b &$\mathrm{\log ({M_{HI}/M_{*,disc}})}$ &$\mathrm{(g-r)_{disc}}$ &$\mathrm{M_{*,disc}}$ &B &-0.70$\pm$0.02 &[-0.73,-0.66] &5.43e-126 &-0.74$\pm$0.02 &[-0.77,-0.70] &1.44e-147 &-0.49$\pm$0.03 &[-0.54,-0.43] &5.96e-52\\
4-a &$\mathrm{\log ({M_{HI}/M_{*,bulge}})}$ &$\mathrm{(g-r)_{bulge}}$ &$\mathrm{M_{*,bulge}}$ &Sersic $n<2.5$ &-0.74$\pm$0.02 &[-0.78,-0.70] &3.17e-99 &-0.75$\pm$0.02 &[-0.78,-0.71] &9.45e-103 &-0.26$\pm$0.04 &[-0.34,-0.18] &2.37e-10\\
4-b &$\mathrm{\log ({M_{HI}/M_{*,bulge}})}$ &$\mathrm{(g-r)_{bulge}}$ &$\mathrm{M_{*,bulge}}$ &Sersic $n>2.5$ &-0.73$\pm$0.03 &[-0.78,-0.67] &2.29e-50 &-0.56$\pm$0.05 &[-0.63,-0.48] &8.84e-26 &-0.30$\pm$0.05 &[-0.40,-0.19] &1.30e-07\\
4-c &$\mathrm{\log ({M_{HI}/M_{*,bulge}})}$ &$\mathrm{(g-r)_{bulge}}$ &$\mathrm{M_{*,bulge}}$ &barred galaxies &-0.81$\pm$0.02 &[-0.85,-0.76] &1.34e-55 &-0.76$\pm$0.03 &[-0.81,-0.71] &8.46e-47 &-0.18$\pm$0.07 &[-0.30,-0.05] &5.82e-03\\
4-d &$\mathrm{\log ({M_{HI}/M_{*,bulge}})}$ &$\mathrm{(g-r)_{bulge}}$ &$\mathrm{M_{*,bulge}}$ &interaction$^a$ &-0.77$\pm$0.02 &[-0.81,-0.72] &4.64e-61 &-0.75$\pm$0.03 &[-0.79,-0.69] &2.94e-56 &-0.31$\pm$0.05 &[-0.41,-0.21] &1.53e-08\\
4-e &$\mathrm{\log ({M_{HI}/M_{*,bulge}})}$ &$\mathrm{(g-r)_{bulge}}$ &$\mathrm{M_{*,bulge}}$ &interaction$^b$ &-0.69$\pm$0.07 &[-0.80,-0.54] &3.93e-11 &-0.67$\pm$0.08 &[-0.78,-0.52] &1.66e-10 &-0.26$\pm$0.12 &[-0.46,-0.02] &3.17e-02\\
5-a &$\mathrm{M_{*,bulge}}$ &$\mathrm{(g-r)_{bulge}}$ &$\mathrm{M_{*,total}}$ &B &0.79$\pm$0.01 &[0.76,0.81] &8.54e-182 &0.79$\pm$0.01 &[0.76,0.81] &1.04e-180 &0.75$\pm$0.02 &[0.72,0.77] &1.23e-153\\
5-b &$\mathrm{M_{*,bulge}}$ &$\mathrm{(g-r)_{bulge}}$ &$\mathrm{Metallicity}$ &Metal &0.77$\pm$0.02 &[0.73,0.80] &4.82e-95 &0.79$\pm$0.02 &[0.75,0.82] &2.46e-104 &0.71$\pm$0.02 &[0.66,0.75] &4.15e-76\\
5-c &$\mathrm{M_{*,bulge}}$ &$\mathrm{(g-r)_{bulge}}$ &$\mathrm{E(B-V)}$ &EBV &0.79$\pm$0.01 &[0.76,0.81] &3.77e-163 &0.79$\pm$0.01 &[0.76,0.82] &1.17e-167 &0.77$\pm$0.01 &[0.74,0.80] &7.63e-154\\
6-a &$\mathrm{\log ({M_{HI}/M_{*,disc}})}$ &$\mathrm{(g-r)_{disc}}$ &B/T &B &-0.70$\pm$0.02 &[-0.73,-0.66] &5.43e-126 &-0.74$\pm$0.02 &[-0.77,-0.70] &1.44e-147 &-0.63$\pm$0.02 &[-0.67,-0.59] &2.18e-97\\
6-b &$\mathrm{\log ({M_{HI}/M_{*,disc}})}$ &$\mathrm{(g-r)_{disc}}$ &$\mathrm{M_{*,bulge}}$ &B &-0.70$\pm$0.02 &[-0.73,-0.66] &5.43e-126 &-0.74$\pm$0.02 &[-0.77,-0.70] &1.44e-147 &-0.73$\pm$0.01 &[-0.76,-0.70] &7.55e-143\\
6-c &$\mathrm{\log ({M_{HI}/M_{*,disc}})}$ &$\mathrm{(g-r)_{disc}}$ &$ \mathrm{\mu_{*,bulge}} $ &B &-0.70$\pm$0.02 &[-0.73,-0.66] &5.43e-126 &-0.74$\pm$0.02 &[-0.77,-0.70] &1.44e-147 &-0.70$\pm$0.02 &[-0.73,-0.66] &2.62e-126\\
6-d &$\mathrm{\log ({M_{HI}/M_{*,disc}})}$ &$\mathrm{(g-r)_{disc}}$ &Sersic n &B &-0.70$\pm$0.02 &[-0.73,-0.66] &5.43e-126 &-0.74$\pm$0.02 &[-0.77,-0.70] &1.44e-147 &-0.70$\pm$0.02 &[-0.73,-0.66] &2.20e-126\\
7-a &$\mathrm{(g-r)_{disc}}$ &$\mathrm{(g-r)_{bulge}}$ &$\mathrm{M_{*,bulge}}$ &B &-0.19$\pm$0.03 &[-0.25,-0.12] &3.00e-08 &-0.09$\pm$0.03 &[-0.16,-0.02] &7.88e-03 &-0.31$\pm$0.03 &[-0.37,-0.24] &4.20e-20\\
7-b &$\mathrm{(g-r)_{disc}}$ &$\mathrm{(g-r)_{bulge}}$ &$\mathrm{M_{*,bulge}}$ &Sersic $n<2.5$ &-0.23$\pm$0.04 &[-0.31,-0.15] &3.93e-08 &-0.11$\pm$0.04 &[-0.19,-0.03] &1.02e-02 &-0.34$\pm$0.04 &[-0.41,-0.26] &1.85e-16\\
7-c &$\mathrm{(g-r)_{disc}}$ &$\mathrm{(g-r)_{bulge}}$ &$\mathrm{M_{*,bulge}}$ &Sersic $n>2.5$ &-0.23$\pm$0.06 &[-0.33,-0.12] &7.91e-05 &-0.24$\pm$0.06 &[-0.35,-0.13] &2.64e-05 &-0.26$\pm$0.05 &[-0.36,-0.15] &9.00e-06\\
9-a &B/T &$\mathrm{\log ({M_{HI}/M_{*,total}})}$ &$\mathrm{M_{*,total}}$ &B &-0.15$\pm$0.03 &[-0.21,-0.08] &1.28e-05 &-0.15$\pm$0.03 &[-0.21,-0.08] &1.80e-05 &0.02$\pm$0.03 &[-0.04,0.09] &5.12e-01\\
9-b &$ \mathrm{\mu_{*,total}} $ &$\mathrm{\log ({M_{HI}/M_{*,total}})}$ &$\mathrm{M_{*,total}}$ &B &-0.60$\pm$0.02 &[-0.64,-0.56] &1.92e-85 &-0.60$\pm$0.02 &[-0.64,-0.56] &4.54e-86 &-0.45$\pm$0.03 &[-0.50,-0.40] &1.56e-44\\
9-c &$ \mathrm{\mu_{*,disc}} $ &$\mathrm{\log ({M_{HI}/M_{*,disc}})}$ &$\mathrm{M_{*,disc}}$ &B &-0.71$\pm$0.02 &[-0.74,-0.68] &7.59e-134 &-0.72$\pm$0.02 &[-0.75,-0.68] &4.41e-137 &-0.52$\pm$0.03 &[-0.57,-0.47] &1.63e-61\\

\hline
\end{tabular}
\begin{flushleft}
 $^a$galaxy with close neighbour\\
$^b$galaxy lie above G-A sequence\\
\end{flushleft}
\end{table}
\end{landscape}




\bibliographystyle{mnras}
\bibliography{2019} 








\bsp	
\label{lastpage}
\end{document}